\documentclass[article,twocolumn]{revtex4-2}
\usepackage{epsfig,amssymb,amsmath,amsthm,amsfonts,amsbsy,mathrsfs,pifont}
\usepackage{graphicx}
\usepackage{booktabs}
\usepackage{diagbox}
\usepackage{xcolor,hyperref}

\usepackage{tikz}
\usepackage{arydshln}
\usepackage{tabularray}

\begin{document}


\title{Separating Energy and Entropy Contributions to the Hexatic-Liquid Transitions in Two-Dimensional Repulsive Systems}
\author{Yan-Wei Li}
\email{yanweili@bit.edu.cn}
\author{Rui Ding}
\author{Wen-Hao Ma}
\affiliation{Key Laboratory of Advanced Optoelectronic Quantum Architecture and Measurement (MOE), School of Physics, Beijing Institute of Technology, Beijing, 100081, China}
\date{\today}
             
\begin{abstract}
Over the past decades, research on two-dimensional melting has established that both first-order and continuous hexatic-liquid transitions can occur, influenced by various factors in the potential energy and system details. The fundamental thermodynamic origins of this sensitivity remains elusive. Here, by decomposing the Helmholtz free energy across three representative repulsive systems, we reveal a universal competition between energy and entropy that dictates the melting pathway. The energetic contribution consistently imparts convexity to the free energy, whereas entropy imparts concavity. A first-order transition occurs when concave entropy dominates; otherwise, the transition is continuous. Further decomposition shows that vibrational entropy drives the concave total entropic curvature, while the configurational entropy's curvature switches from convex (first-order) to concave (continuous), mirroring defect proliferation measured by Shannon entropy. The convexity of the energy is dominated by the inherent potential, with minimal vibrational influence. Finally, we predict and verify that the first-order transition becomes continuous at zero temperature, where entropic effects vanish. Our work establishes the curvature of different thermodynamic quantities as a fundamental principle for understanding the nature of two-dimensional melting.
\end{abstract}

\maketitle

Two-dimensional (2d) systems exhibit complex physics of phase transitions, which has significantly contributed to recent advances in statistical mechanics. Different from three-dimensional solids, which typically undergo a first-order transition directly from solid to liquid without any intermediate phase, the melting process of two-dimensional solids has remained a longstanding mystery. The well-known Mermin-Wagner theorem excludes the possibility of true long-range translational order in 2d systems with short-range continuous interactions due to the impact of long-wavelength fluctuations~\cite{Mermin}. Later, the Kosterlitz-Thouless-Halperin-Nelson-Young (KTHNY) theory introduced a new class of 2d solids characterized by long-range bond-orientational order and quasi-long-range translational order, as well as the existence of an intermediate hexatic phase between solid and liquid states~\cite{KT,HN,Y}. According to the KTHNY theory, the melting of a 2d solid into a liquid proceeds via a two-step transition through an intermediate hexatic phase~\cite{KT,HN,Y}. This hexatic phase exhibits quasi-long-range bond-orientational order but only short-range translational order. Both transitions are mediated by the accumulation of topological defects. The solid-to-hexatic transition occurs when bound dislocations dissociate into free dislocations. Subsequently, the hexatic-to-liquid transition involves the dissociation of these dislocations into free disclinations. While the KTHNY theory posits that the transitions are typically continuous, it does not exclude the possibility of a first-order transition.

Numerical and experimental verification of the 2d melting mechanism has been challenging due to the large hexatic correlation length and the prolonged equilibration time of the hexatic phase. Only recently has the melting behavior become clearer in the simplest hard disc system, where it was found that the hexatic-liquid transition is actually first-order~\cite{Krauth2011,Experiment_harddisc}, in contrast to the long‑assumed picture of either a two‑step continuous scenario or a one‑step first‑order scenario.

Recent studies have demonstrated that even subtle changes in particle interactions can significantly influence the nature of the hexatic-liquid transition~\cite{Krauth2015,ningxu,John_Russo,Glotzer,Tanaka,Cell,LJ_pd,Massimo_2020PRLmelting,YW_MP_prl2023,zhuwang_NC2024,ZirenPRX,deform_PRL2020}. For example, in soft-core systems like those with Hertzian or Gaussian interactions, the liquid-hexatic transition and an additional reentrant transition can occur at different densities~\cite{ningxu}. At low densities, the transition is typically first-order, while at high densities, it becomes continuous\cite{ningxu}. In systems with an inverse power-law potential $U(r) = \epsilon (\sigma / r)^n$, increasing the exponent $n$, which enhances the interaction stiffness, can shift the hexatic-liquid transition from continuous to first-order~\cite{Krauth2015}. Furthermore, the nature of the hexatic-liquid transition is highly sensitive to factors such as the potential's attractive component~\cite{ZirenPRX,Massimo_2020PRLmelting,LJ_pd,dudi_softmatter}, particle shape~\cite{Glotzer,YW_MP_prl2023,zhuwang_NC2024}, deformability~\cite{Tanaka,Cell,deform_PRL2020}, polydispersity~\cite{John_Russo,Ran2dmelt,Guo2021}, and so on~\cite{Krauth_NC2018,pinned_particle,qi_sm2015,Toledano_PRB,activemelting_prl,Cugliandolo_prl, Bui_Science,zhuwang_NC2024}. These observations raise key questions about what fundamentally affects the nature of the hexatic-liquid transition and why this transition is so sensitive to the various details of the interaction potential.

In this work, we revisit three representative repulsive systems—the soft-core Hertzian and Gaussian models, and the hard-repulsive inverse power-law systems—to explore the thermodynamic properties during the hexatic-liquid transition. We find that the Helmholtz free energy is concave for a first-order transition but convex for a continuous one, consistent with the presence (or absence) of a Mayer–Wood loop in the equation of state. By decomposing the free energy into energetic and entropic components, we observe that concavity originates from the entropic part, while the energy is convex. Further decomposition of the entropy reveals that for a first-order transition, the concave entropy results solely from a concave vibrational contribution, while the configurational contribution is convex. In contrast, for a continuous transition, both entropic contributions are concave. The dependence of configurational entropy curvature on the melting mechanism is closely linked to defects. We quantify this by defining Shannon entropies based on different defect measures, finding that their curvature aligns with that of the configurational entropy. The convexity of the energy mainly arises from the inherent energy, with only a minor contribution from the vibrational component. The interplay between these contributions thus determines the nature of the hexatic-liquid transition. We further predict that the first-order hexatic-liquid transition may become continuous at zero temperature due to the vanishing of entropic contributions. This prediction is supported by the disappearance of the Mayer–Wood loop in the inherent equation of state and the evolution of the local-area distribution from bimodal to unimodal upon minimizing coexisting configurations. Our findings are general across the 2d repulsive models studied; however, in three dimensions (3d), the curvature of energy and entropy can change sign within the first-order solid-liquid transition region.


\section*{Methods}

\subsection*{Model systems}

We study systems with Hertzian, inverse power-law, and Gaussian interactions in 2d, and also a 3d Gaussian system. For the Hertzian interaction, the potential is given by 
\[
U\left(r_{i j}\right) = \frac{\epsilon}{2.5} \left(1-\frac{r_{i j}}{\sigma}\right)^{2.5} \Theta\left(1-\frac{r_{i j}}{\sigma}\right),
\]
where $\epsilon$ is the energy scale, $\sigma$ the particle diameter, $r_{ij}$ the separation between particles $i$ and $j$, and $\Theta$ the Heaviside step function. We fix the temperature at $T=0.003$ and examine two density ranges: a low-density regime $\rho = 1.366\!-\!1.340$ (the HertL system; area per particle $v=1/\rho\simeq 0.7321\!-\!0.7463$) and a high-density regime $\rho = 1.980\!-\!2.020$ (the HertH system; $v\simeq 0.5051\!-\!0.4950$). Melting in these ranges proceeds via a first-order and a continuous hexatic--liquid transition, respectively, as reported previously~\cite{ningxu}.

For the inverse power-law potential, we use  
\[
U(r_{i j}) = \epsilon \left(\frac{\sigma}{r_{i j}}\right)^n,
\]
cut and shifted at $r_c=2.5$ to ensure continuity of the potential at the cutoff. We consider exponents $n=6$ (IPL6) and $n=64$ (IPL64) at temperature $T=1.0$. The IPL6 system exhibits a continuous hexatic--liquid transition, whereas the IPL64 system shows a first-order transition, consistent with earlier results~\cite{Krauth2015}.

The Gaussian interaction is defined as  
\[
U\left(r_{i j}\right) = \epsilon \exp\!\left(-r_{i j}^{2} / \sigma^{2}\right),
\]
cut and shifted the potential at $r_c=4.0$. In 2d (Gauss2d), we fix $T=0.0018$ and examine the density range $\rho = 0.836\!-\!0.887$ ($v\simeq 1.196\!-\!1.127$), where melting is known to be continuous~\cite{ningxu}. In three dimensions (Gauss3d), we set $T = 0.001$ and study the range $\rho = 0.066\!-\!0.06$, over which a face-centered cubic crystal melts into a liquid via a first-order transition, as reported in ref.~\cite{Gauss3d}.

All parameters ($\epsilon$, $\sigma$ and $r_{ij}$) have the same meaning across the three models. We employ reduced units with length, energy, and mass expressed in units of $\sigma$, $\epsilon$, and $m$, respectively. Simulations are performed with $N=318^2$ particles in 2d and approximately one million particles in 3d, using a particle mass $m=1$ and periodic boundary conditions. Temperature is controlled via a Nos\'{e}-Hoover thermostat~\cite{Allen_book}. The molecular dynamics runs are carried out with the GPU-accelerated GALAMOST package~\cite{Galamost}, and energy minimizations are performed using LAMMPS~\cite{Lammps}.

\subsection*{Vibrational density of states and the vibrational entropy}

The vibrational density of state, $g(\omega)$ is computed by performing the Fourier transform of the velocity autocorrelation function~\cite{Shintani_NM,Hu_NP}, 
\[
g(\omega)=\int \frac{\mathrm{d} t}{2 \pi} \exp (i \omega t) \frac{\left\langle\sum \sqrt{m_i} v_i(t) \cdot \sqrt{m_i} v_i(0)\right\rangle}{\left\langle\sum \sqrt{m_i} v_i(0) \cdot \sqrt{m_i} v_i(0)\right\rangle},
\]
where $v_i(t)$ is the velocity of particle $i$ at time $t$ and $m_i=1$ is the mass of particle $i$. The vibrational entropy is calculated from $g(\omega)$\cite{Smith_NP, Allen_PRB}, given as, 

\[
S_{\rm vib}=2 k_B \int_0^{\infty} g(e)[(\mathrm{n}(e)+1) \ln (\mathrm{n}(e)+1)-\mathrm{n}(e) \ln (\mathrm{n}(e))] d e,
\]
where, $k_B$ is the Boltzmann constant, $\mathrm{n}(e)=(\mathrm{exp}[e/(k_BT)]-1)^{-1}$ is the phonon occupancy and $e=\omega\hbar$ is the phonon energy.

\subsection*{Definition of Shanoon entropy of defects}
In addition to the Shannon entropy of the hexatic order parameter $S_6$, we define Shannon entropies for several other defect-related quantities: the Voronoi area, the binary defect state, and the cluster length of defects.

The Shannon entropy of the Voronoi area is defined as
\[
S_v = -\sum_{v} p(v) \ln [p(v)],
\]
where \( p(v) \) is the probability that a particle’s Voronoi area equals  v . The Shannon entropy of the binary defect state is
\[
S_{\mathrm{def}} = -\sum_{k\in\{0,1\}} p(k) \ln [p(k)],
\]
with \( p(k) \) the probability that a particle’s defect state is  k . Finally, the Shannon entropy of the defect‑cluster length is defined as
\[
S_{\mathrm{clus}} = -\sum_{n} p(n) \ln [p(n)],
\]
where \( p(n) \) is the probability of finding a defect cluster composed of  n  defective particles.

\subsection*{Local area}

We characterize the coexistence state using the probability distribution of the local area. For each particle located at $\vec{r}_i$, we define a local area per particle as 
\[ 
v_{\rm loc}(\vec{r}_i) = \frac{\pi R_c^2}{\sum_{j=1}^N H\!\left(R_c - |\vec{r}_i - \vec{r}_j|\right)} ,
\] 
where $H$ is the Heaviside step function and $R_c = 40$. The results are robust with respect to the choice of $R_c$ within a reasonable range.

\begin{figure}[!t]
\centering
\includegraphics[angle=0,width=0.5\textwidth]{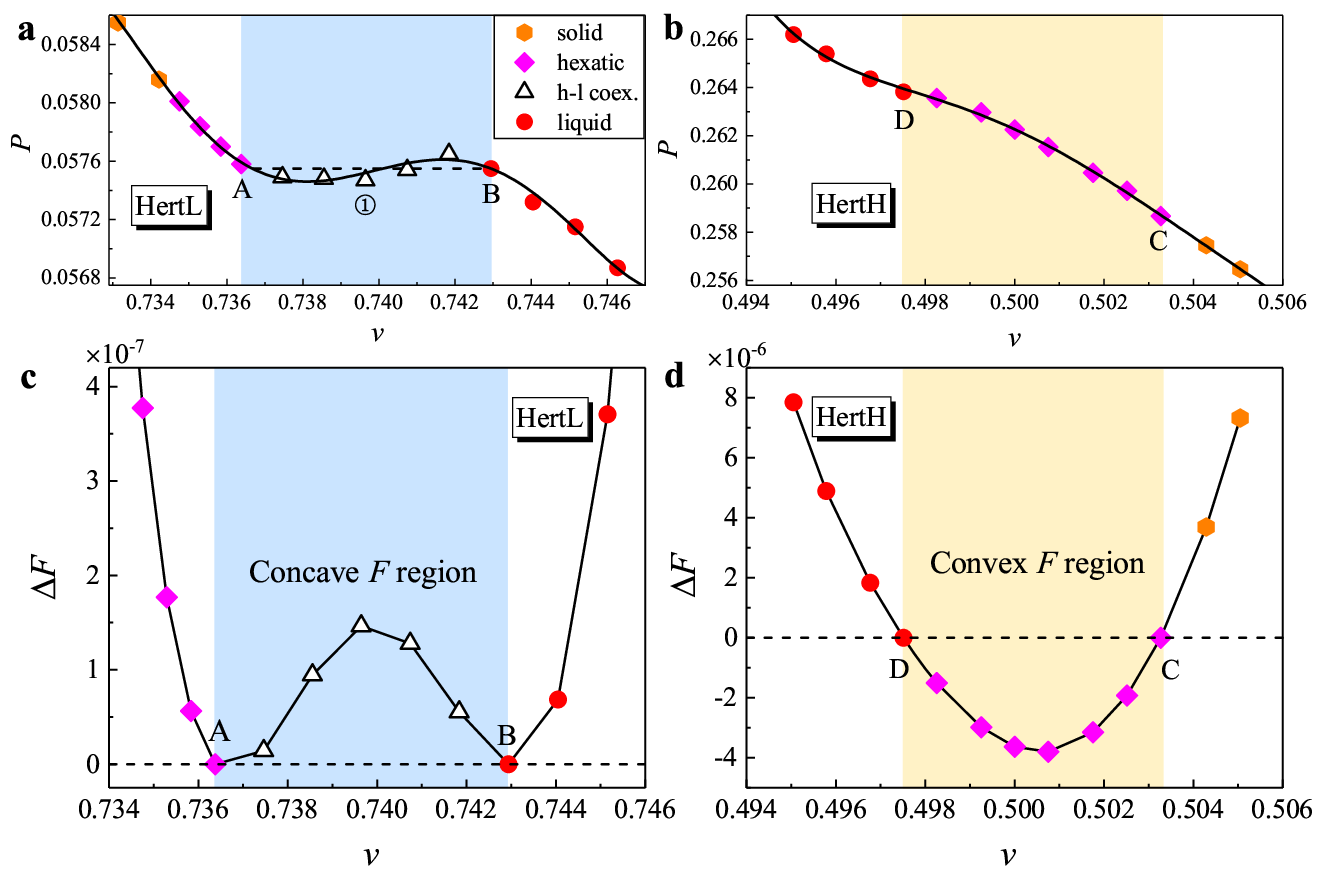}
\caption{
\textbf{Isothermal equation of state and Helmholtz free energy curvature.}
\textbf{a}, \textbf{b} Equilibrium isothermal equation of state $P(v)$ for the Hertzian system at \textbf{a} low and \textbf{b} high density. Black solid lines are fifth-order polynomial fits. The horizontal dashed line in \textbf{a} indicates the Maxwell construction. \textbf{c}, \textbf{d} Corresponding difference in Helmholtz free energy, $\Delta F = F - F_L$, for \textbf{c} the low-density HertL system and \textbf{d} the high-density HertH system. Here, $F$ is the Helmholtz free energy (see Supplementary Fig. S2), and $F_L$ is a linear reference function connecting the free energy at two boundary states: between $F(v_A)$ and $F(v_B)$ for HertL system, and between $F(v_C)$ and $F(v_D)$ for HertH system. The sign of $\Delta F$ quantifies the curvature of $F(v)$: $\Delta F > 0$ (shaded light blue in \textbf{c}) indicates concave curvature, characteristic of the first-order hexatic–liquid transition; $\Delta F < 0$ (shaded yellow in \textbf{d}) indicates convex curvature, characteristic of the continuous hexatic–liquid transition. Symbols in all panels correspond to different states, as shown in the legend of panel \textbf{a}.
}
\label{fig:eos}
\end{figure}

\section{Results}

In the main text, we primarily present results for the Hertzian model at a low density (HertL) and a high density (HertH), where the hexatic-liquid transition is first-order and continuous, respectively. Consistent results are illustrated for a system with an inverse power-law interaction in Supplementary Fig. S4. Here, the nature of the hexatic-liquid transition depends on the exponent n (see Methods for details), with the n=64 (IPL64) system exhibiting a first-order transition and the n=6 (IPL6) system a continuous one. Additionally, we present results for the Gaussian model in both two and three dimensions in Supplementary Figs. S4 and S5. These show consistent behavior in 2d but reveal a different scenario in 3d. The study in these different systems allows us to construct a general picture of the thermodynamic properties during hexatic-liquid transitions in 2d repulsive systems and to distinguish it from the 3d case, where the hexatic phase is absent.

\subsection{The concavity and convexity of free energy during the hexatic-liquid transitions}

\begin{figure*}[!tb]
\centering
\includegraphics[angle=0,width=0.8\textwidth]{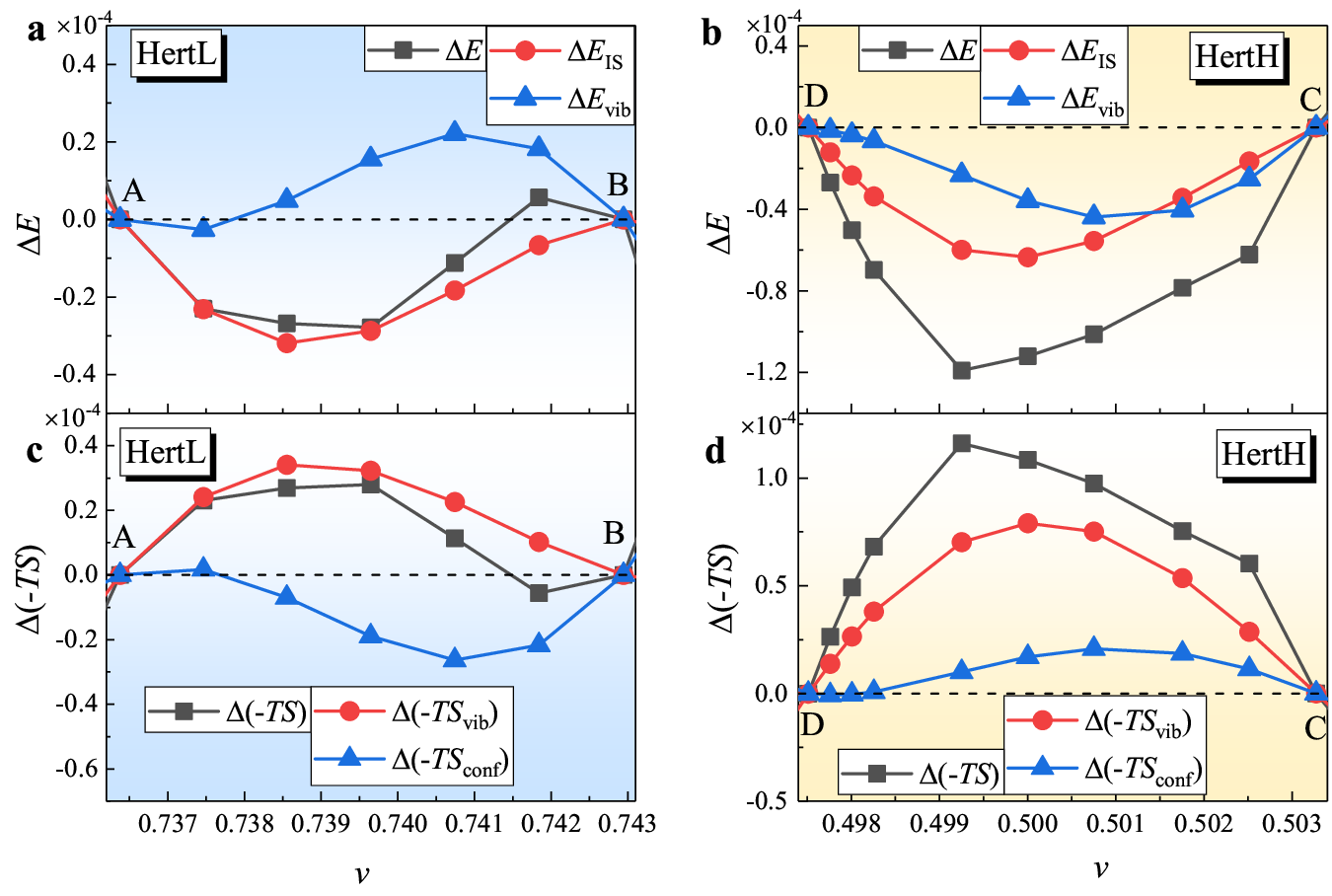}
\caption{\textbf{Decomposed various energy and entropy contributions to curvature.} \textbf{a}, \textbf{b} Difference in total energy $\Delta E$ (black squares), inherent energy $\Delta E_{\rm IS}$ (red circles), and vibrational energy $\Delta E_{\rm vib}$ (blue triangles) for \textbf{a} the HertL and \textbf{b} the HertH systems. Analogous to $\Delta F$ in Figs.~\ref{fig:eos}\textbf{c} and \ref{fig:eos}\textbf{d}, the sign of these functions determines the local curvature (concavity or convexity) of the corresponding function. \textbf{c}, \textbf{d} Difference in total entropic $\Delta (-TS)$ (black squares), vibrational entropy $\Delta (-TS_{\rm vib})$ (red circles), and configurational entropy $\Delta (-TS_{\rm conf})$ (blue triangles), for the two systems. The data in all panels are shown only for the transition regions (shaded in Fig.~\ref{fig:eos}).
}
\label{fig:enHer}
\end{figure*}

Figures~\ref{fig:eos}a and \ref{fig:eos}b show the equations of state, $P(v)$, i.e., the pressure as a function of the area per particle $v = 1/\rho$, where $\rho$ is the number density, for the HertL and HertH systems, respectively. Different states are indicated by different symbols, with details of the determination of states provided in Supplementary Fig. S1. Consistent with the findings in ref.~\cite{ningxu}, Hertzian particles undergo a first-order hexatic-liquid transition at low density, resulting in the Mayer-Wood loop observed in the equation of state in Fig.~\ref{fig:eos}a~\cite{Mayer_wood}, which characterizes the coexistence of the hexatic and liquid phases. In contrast, at high density, the pressure decreases monotonically with increasing $v$, with no pressure loop (Fig.~\ref{fig:eos}b), suggesting a continuous hexatic-liquid transition.

We further compute the Helmholtz free energy per particle via thermodynamic integration $F(v)=F_0-\int_{v_0}^v P(v)dv$, where $P(v)$ is fitted with a fifth-order polynomial, as shown by the black solid lines in Figs.~\ref{fig:eos}a and \ref{fig:eos}b. Here $v_0$ is taken as the smallest $v$ studied, and $F_0$ is chosen such that $F(v)=0$ at the largest $v$ investigated. The resulting $F(v)$ is plotted as symbols in Supplementary Fig. S2. To investigate whether $F(v)$ is concave in the hexatic-liquid transition region, we first define a linear function $F_L(v)$, which represents the line connecting $F(v_A)$ and $F(v_B)$, where $v_A$ and $v_B$ correspond to the area per particle at states $A$ and $B$, respectively, marked in both Fig.~\ref{fig:eos}(a) and Fig.~\ref{fig:eos}(c). These state points lie on the two sides of the coexistence boundary, as determined by the Maxwell construction. We then compute the difference between $F(v)$ and $F_L(v)$ as $\Delta F(v) = F(v) - F_L(v)$. The sign of $\Delta F(v)$ indicates the local curvature of $F(v)$: $\Delta F(v) > 0$ signifies a concave region, while $\Delta F(v) < 0$ indicates a convex one. Figure~\ref{fig:eos}(c) illustrates $\Delta F(v)$ for the HertL system, where we observe $\Delta F(v) > 0$ in the hexatic-liquid coexistence region (marked as a light blue shaded area), as expected. This result is further highlighted by comparing $\Delta F(v)$ for the HertH system. In the HertH system, $F_L(v)$ is a linear function connecting $F(v_C)$ and $F(v_D)$, where state point C represents the liquid boundary nearest to the hexatic phase and state point D represents the hexatic boundary nearest to the solid phase, as indicated in Figs.~\ref{fig:eos}b and \ref{fig:eos}d. In contrast, for the HertH system, we observe negative values of $\Delta F(v)$ in the hexatic and hexatic-liquid transition regions (marked as a light yellow shaded area), suggesting that $F(v)$ is convex during the continuous hexatic-liquid transition. The concave $F(v)$ for the HertL is consistently characterized by negative values of the second derivative of $F(v)$ where $F(v)$ is fitted by a 5th-order polynomial, as presented in Supplementary Fig. S2. Similarly, we find positive values of the second derivative of $F(v)$, showing its convex nature for the HertH system. Although the concave or convex nature of $F(v)$ does not depend on studying either $\Delta F(v)$ or the second derivative of $F(v)$, we stick to the former, as this method does not require any fitting of $F(v)$.

\subsection{Separating the energy and entropy contributions during the transitions}

\begin{figure}[!!t]
\centering
\includegraphics[angle=0,width=0.45\textwidth]{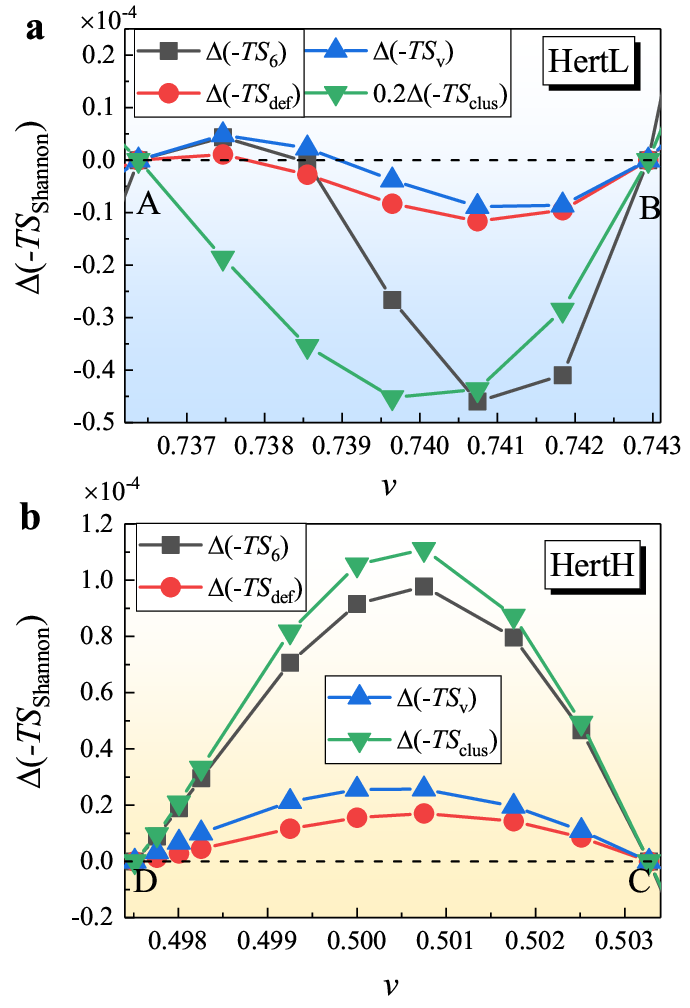}
\caption{\textbf{Curvature of various Shannon entropies defined from different defect quantifications.} Difference in Shannon entropy $\Delta (-TS_{\rm Shannon})$, for \textbf{a} HertL and \textbf{b} HertH systems. The Shannon entropy is calculated from four different measures: the hexatic order parameter $\Delta (-TS_6)$ (black squares), a binary defect state $\Delta (-TS_{\rm def})$ (red circles), the Voronoi area $\Delta (-TS_v)$ (blue triangles), and the cluster length of defects $\Delta (-TS_{\rm clus})$ (green triangles). Analogous to $\Delta F$ in Figs.~\ref{fig:eos}\textbf{c} and \ref{fig:eos}\textbf{d}, the sign of $\Delta (-TS_{\rm Shannon})$ determines the local curvature (concavity or convexity) of the corresponding Shannon entropic contribution. Data are shown only for the transition regions (shaded in Fig.~\ref{fig:eos}).
}
\label{fig:Shannon}
\end{figure}

To investigate the origin of the concavity in the Helmholtz free energy, we decompose the free energy into its energetic and entropic components, expressed as $ F(v) = E(v) - TS(v)$. The total energy, $ E(v)$, is readily obtained from simulations. Analogous to $\Delta F(v)$, we calculate the difference between $ E(v)$ and $ E_L(v) $ as $\Delta E(v) = E(v) - E_L(v)$, where $E_L(v)$ represents a linear interpolation between $ E(v_A)$ and $ E(v_B)$ for the HertL system, and between $ E(v_C) $ and $ E(v_D) $ for the HertH system. We illustrate $\Delta E$ as black squares for both the HertL and HertH systems in Figs.~\ref{fig:enHer}a and \ref{fig:enHer}b, respectively. For both systems, $\Delta E$ is mostly negative within the relevant transition regions: specifically, in the hexatic-liquid coexistence region (light blue shading in Fig.~\ref{fig:eos}a) for HertL, and in the hexatic and hexatic–liquid transition regions (light yellow shading in Fig.~\ref{fig:eos}b) for HertH. This indicates that the energy component consistently favors a convex contribution to the free energy and therefore cannot drive a first-order transition in these systems. 

According to the potential energy landscape formalism, the total energy of a system in thermal equilibrium can be decomposed into a contribution from the energy of the minimized state—known as the inherent structure—and a contribution from the vibrational energy relative to the inherent structure, expressed as $E=E_{\rm IS} + E_{\rm vib}$~\cite{Sciortino_2005,Sciortino2016,gangsunPRL2018}. To investigate the origin of the convex energy contribution, we minimize the thermal equilibrium configurations at different densities to their inherent structures using the Fast Inertial Relaxation Engine (FIRE) algorithm~\cite{Fire}. This yields the inherent energy $E_{\rm IS}$ and, consequently, the vibrational energy $E_{\rm vib}=E-E_{\rm IS}$. 
Following the same method applied to free energy and the total energy, the curvature of $E_{\rm IS}$ and $E_{\rm vib}$ is quantified by $\Delta E_{\rm IS}$ and $\Delta E_{\rm vib}$, respectively. For the HertL system (Fig.~\ref{fig:enHer}a), we find negative $\Delta E_{\rm IS}$ but positive $\Delta E_{\rm vib}$, indicating that the convexity of the total energy $E$ is dominated by the convex inherent energy $E_{\rm IS}$ and is partially offset by the concave vibrational energy $E_{\rm vib}$. In contrast, for the HertH system (Fig.~\ref{fig:enHer}b), both $E_{\rm IS}$ and $E_{\rm vib}$ are convex, resulting in a definitively convex total energy. These findings underscore an important correspondence between curvature of thermal energy and its inherent energy, consistent with our recent finding that, although the defect density differs markedly, the phase of the thermal system is identical to that of the inherent configuration across a two-step continuous melting transition~\cite{MaPRE}.

We next consider the entropy, which should be the principal driver of the first-order transition, as the energy contribution has been ruled out. The entropic term is derived from the relation $-TS = F - E$. We quantify its curvature by calculating $\Delta(-TS)$ using the same method as before. The resulting $\Delta(-TS)$ values are shown as black squares for the HertL and HertH systems in Figs.~\ref{fig:enHer}c and \ref{fig:enHer}d, respectively. For both systems, we observe $\Delta(-TS)>0$ in the considered regions, indicating that the concavity of the Helmholtz free energy and thus the first-order transition of the system with Hertzian interactions arises from the concavity of the entropic component. 

The entropy can be further decomposed into vibrational and configurational contributions, i.e., $S= S_{\rm vib}+ S_{\rm conf}$. This decomposition has been widely assessed in both crystalline solids and amorphous systems~\cite{Huang,Johari2000,Smith_NP}, but not in the hexatic-liquid transitions in two dimensions. The vibrational entropy $S_{\rm vib}$ is evaluated via the density of states (DOS), which is obtained from the Fourier transform of velocity correlation functions, as detailed in the Methods. Typical results for the DOS at different states and the entropies $S$, $S_{\rm vib}$, and $S_{\rm conf}$ for the HertL system are illustrated in Supplementary Fig. S3. The curvature of the entropies, quantified by $\Delta (-TS)$, $\Delta (-TS_{\rm vib})$, and $\Delta (-TS_{\rm conf})$, is presented in Figs.~\ref{fig:enHer}c and \ref{fig:enHer}d. For the HertL system, $\Delta (-TS_{\rm vib})$ is positive while $\Delta (-TS_{\rm conf})$ is negative, indicating a concave vibrational contribution and a convex configurational contribution to the entropy. The vibrational term dominates, yielding a net concave total entropy, which in turn produces the concave free energy responsible for the first-order hexatic-liquid transition. We therefore conclude that the first-order hexatic-liquid transition in the HertL system is driven exclusively by the concave vibrational entropy, as all other free energy components—the total energy and configurational entropy—are convex. In contrast, for the HertH system, both the vibrational and configurational entropic contributions are concave, resulting in an even more concave total entropy. Nevertheless, this concave entropy is overcome by the convex energy, leading to a convex free energy and thus a continuous hexatic–liquid transition in the HertH system.


\subsection{Shannon entropy of defects}

Since entropy is intrinsically linked to disorder, its increase (Supplementary Fig. S3) as a solid melts coincides with the accumulation of defects. To understand how defects influence the curvature of entropy, we next quantify defects using four measures: the hexatic order parameter, a binary defect state, the Voronoi area, and the cluster length of defects. Here, a particle's binary defect state is assigned a value of 0 if it has 6 nearest neighbors and 1 otherwise. The first three quantities characterize defects at the level of individual particles. The fourth, the cluster length, accounts for the widely observed phenomenon that defective particles can agglomerate into large clusters~\cite{digregorio2019clustering,LJ_pd,Krauth2015,Glotzer}—a feature not considered in the original KTHNY theory~\cite{KT,HN,Y}. We compute the Shannon entropy for these four quantities. For instance, the Shannon entropy of the hexatic order parameter is defined as $S_6 = -\sum_{\psi_{6}=0}^{1} p(\psi_{6}) \ln{[p(\psi_{6})]}$, where $p(\psi_{6})$ is the probability that a particle’s hexatic order parameter equals $\psi_{6}$. The Shannon entropies of the Voronoi area, the binary defect state, and the cluster length of defects are denoted as $S_v$, $S_{\rm def}$ and $S_{\rm clus}$, respectively; their definitions are provided in the Methods.

\begin{table*}
    \centering
    \setlength{\tabcolsep}{6pt}
    \caption{\textbf{Curvature of thermodynamic quantities across first-order and continuous transitions.} Summary of concave ($+$) and convex ($-$) curvature in the hexatic-liquid transition region for key thermodynamic quantities: Helmholtz free energy $F$, total energy $E$, total entropy $-TS$, vibrational entropy $-TS_{\rm vib}$, configurational entropy $-TS_{\rm conf}$, and the Shannon entropy of defects $-TS_{\rm Shnnon}$. Results are shown for systems with a first-order transition (HertL, iPL64) and for systems with a continuous transition (HertH, IPL6, Gauss). 
    }
    \begin{tabular}{cccccccc}
        \toprule
        \diagbox{System}{Feature} & hexatic-liquid transition & $F$ & $E$ & $-TS$ & $-TS_{\rm vib}$ & $-TS_{\rm conf}$ & $-TS_{\rm Shannon}$ \\
        \midrule
        HertL and IPL64 & first-order & $+$ & $-$ & $+$ & $+$ & $-$ & $-$ \\
        \midrule
        HertH, IPL6 and Gauss2d & continuous  & $-$ & $-$ & $+$ & $+$ & $+$ & $+$ \\
        \bottomrule
    \end{tabular}
    \label{tab:Delta}
\end{table*}

We observe a monotonic increase in the Shannon entropy of defects as the solid melts into the liquid, as exemplified in Supplementary Fig. S3. A notable feature is that the Shannon entropy rises in a manner similar to—and with values close to—the configurational entropy. Both quantities begin to increase already in the solid state, whereas the vibrational entropy starts to grow only near the end of the hexatic region. The curvature of these Shannon entropies is quantified by $\Delta S_6$, $\Delta S_v$, $\Delta S_{\rm def}$, and $\Delta S_{\rm clus}$, illustrated in Fig.~\ref{fig:Shannon}a for the HertL system and in Fig.~\ref{fig:Shannon}b for the HertH system. These curvature measures are negative for the HertL system and positive for the HertH system. This indicates that, independent of the specific defect measure, the Shannon entropy of defects as a function of area per particle $v$ is convex when the hexatic–liquid transition is first-order and concave when the transition is continuous—a pattern that is further corroborated by results from other models presented below. The curvature of the Shannon entropy of defects again aligns closely with that of the configurational entropy, with both exhibiting the same trends across the two systems. This correspondence indicates that defects are directly linked to the configurational entropy, which contributes to the total entropy together with the vibrational entropy—whose curvature, in contrast, is not directly associated with defects.

\subsection{Inherent melting}

\begin{figure}[!!t]
\centering
\includegraphics[angle=0,width=0.4\textwidth]{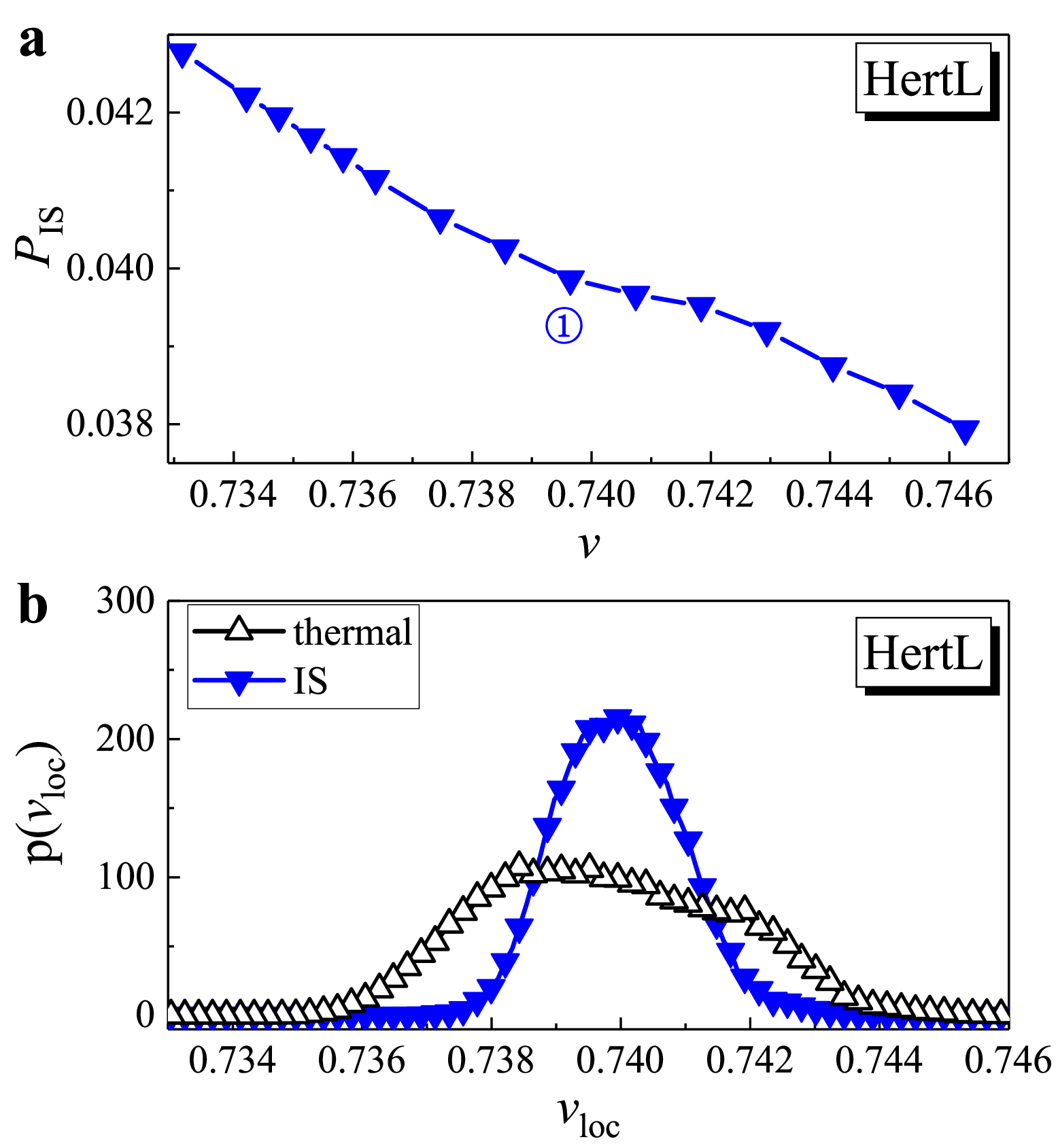}
\caption{\textbf{Melting at zero temperature for the HertL system.}
\textbf{a} Inherent equation of state for the HertL system. \textbf{b} Probability distribution of the local area per particle for the thermal equilibrium configuration (black open triangles) and the inherent structure (blue solid triangles), at state point \ding{172} marked in panel \textbf{a} as well as in Fig.~\ref{fig:eos}\textbf{a}.
}
\label{fig:inherent}
\end{figure}

At zero temperature, all entropic contributions vanish. As discussed above, the first-order hexatic-liquid transition in the HertL system originates from concave entropy. We therefore predict that this first-order transition will be absent at zero temperature. This prediction is consistent with the convex $E_{\rm IS}$ shown in Fig.~\ref{fig:enHer}a. To provide more direct evidence for continuous inherent melting, we present the inherent equation of state in Fig.~\ref{fig:inherent}a. Unlike the isothermal case at thermal equilibrium depicted in Fig.~\ref{fig:eos}a, where a Mayer-Wood loop appears within the hexatic-liquid coexistence region, the pressure loop in this region disappears at zero temperature (see Fig.~\ref{fig:inherent}a).

We further calculate the probability distribution $p(v_{\rm loc})$ of the local area (see Methods) per particle for a specific state point \ding{172} within the hexatic-liquid coexistence region, as marked in Figs.~\ref{fig:eos}a and \ref{fig:inherent}a. The distribution $p(v_{\rm loc})$ for the thermal equilibrium configuration is shown with black triangles, while the inherent structure configuration is represented with blue triangles in Fig.~\ref{fig:inherent}b. Our findings reveal a bimodal distribution of $p(v_{\rm loc})$ for the thermal equilibrium configuration and a unimodal distribution for the inherent structure. This provides clear evidence for the disappearance of the hexatic-liquid coexistence phase and thus the absence of a first-order hexatic-liquid transition at zero temperature. These observations align with findings in two-dimensional cellular systems, where the hexatic-liquid coexistence region diminishes at low temperatures and vanishes at zero temperature~\cite{Cell}.

\subsection{Generality of the findings}

Next, we extend these findings to additional repulsive models to confirm that the described scenario is not unique to the Hertzian system. We consider two systems with an inverse-power-law potential~\cite{Krauth2015}: one with a high exponent ($n = 64$, IPL64) and one with a lower exponent ($n = 6$, IPL6) (see Methods for details). The IPL64 and IPL6 systems exhibit first-order and continuous hexatic-liquid transitions, respectively~\cite{Krauth2015}. We also analyze the soft-repulsive Gaussian model at a relatively high density, where the hexatic-liquid transition is continuous (model details are provided in the Methods).

In Table~\ref{tab:Delta}, we summarize the curvature of key thermodynamic quantities, including the free energy $F$, total energy $E$, entropic contribution $-TS$, and the decomposed configurational and vibrational entropies, as well as the Shannon entropy of various defect measures, for all 2d systems studied. The corresponding data are detailed in Supplementary Fig. S4. For every system, the free energy is concave ($+$) in the transition region when the hexatic-liquid transition is first-order and convex ($-$) when the transition is continuous. This matches our expectation and validates our method for identifying curvature. More intriguingly, regardless of model details or the specific melting scenario, the energy consistently contributes a convex curvature to the free energy, whereas the entropy contributes a concave curvature. A first-order hexatic-liquid transition occurs when the concave entropic contribution dominates; otherwise, the transition is continuous. Decomposing the entropy further reveals that the vibrational entropy contribution is always concave, independent of the melting scenario, while the configurational entropy depends on it: it is convex for a first-order transition and concave for a continuous one. Finally, the curvature of the Shannon entropy of defects aligns closely with that of the configurational entropy for all the investigated systems.

\section{Discussion}

In summary, decomposing the Helmholtz free energy into energetic and entropic contributions across various repulsive systems reveals that their interplay determines the melting scenario. While the entropic component is concave, the energetic component is convex. Consequently, a first-order hexatic-liquid transition occurs when the concave entropic contribution dominates, whereas a continuous transition is observed otherwise. Further decomposition of the entropy into vibrational and configurational parts shows a robustly concave vibrational contribution, while the curvature of the configurational entropy depends on the nature of the hexatic–liquid transition: it is convex for a first-order transition and concave for a continuous transition, and is closely linked to defects. These findings suggest that for a first-order hexatic–liquid transition, the concave vibrational entropy is the sole driver. The convexity of the energy is dominated by the potential energy minima and is only weakly affected by the vibrational energy around those minima. We predict-and demonstrate through the inherent equation of state and the probability distribution of the local area per particle-that the first-order hexatic-liquid transition becomes continuous at zero temperature as the entropic contribution vanishes. Our findings may open a route toward a fundamental understanding of the long-observed yet previously unexplained sensitivity of the hexatic--liquid transition to system potential parameters and specific details~\cite{Krauth2015,ningxu,John_Russo,Glotzer,Tanaka,Cell,LJ_pd,Massimo_2020PRLmelting,YW_MP_prl2023,zhuwang_NC2024,ZirenPRX,deform_PRL2020}. Specifically, there is a delicate balance between the convexity of the energy and the concavity of the entropy, which can be readily altered by system parameters.

The concavity or convexity of physical quantities may be related to the presence of a restoring force. Consider, for example, the harmonic potential $U = \frac{1}{2} k x^2$, which is convex. When a particle moves away from equilibrium, a restoring force pulls it back, corresponding to a minimum in the potential. Conversely, if a force drives the particle away from equilibrium, we would expect a concave potential with a maximum. By analogy, if a restoring force acts on the entropy-making it more convex (or less concave)-it could cause a first-order hexatic-liquid transition to become continuous. In this context, we note that for hard polygons~\cite{Glotzer}, continuous hexatic-liquid transitions occur only for systems with a small number of edges ($<7$), where a strong directional entropic force stabilizes edge-to-edge alignment and promotes locally ordered solid motifs~\cite{Glotzer}. The directional entropic force quantified in ref.~\cite{Glotzer} may thus serve as the restoring force that promotes a continuous transition.

In three dimensions, previous results indicate a first-order transition at zero temperature in Hertzian~\cite{hert3d} and Gaussian~\cite{Gauss3d} systems, albeit not a hexatic–liquid transition. To examine this, we study the melting of a face-centered cubic crystal into a liquid in a Gaussian repulsive system at a low temperature (T=0.001, see Methods for details). We observe a concave Helmholtz free energy, confirming the first-order solid–liquid transition (Supplementary Fig. S5). The decomposed energetic and entropic contributions, however, do not exhibit a uniform curvature in the coexistence region; rather, their curvature changes sign within this region. This behavior differs from the picture seen in 2d and helps rationalize the possible existence of a first-order solid–liquid transition even at zero temperature.

Our findings may inspire further theoretical and experimental studies on the physical origins of different two-dimensional melting scenarios, including those not discussed here, such as first-order solid–liquid transitions in the absence of a hexatic phase~\cite{Glotzer,Tanaka,Cell,Massimo_2020PRLmelting,YW_MP_prl2023}. The systems studied here are all purely repulsive; it would also be interesting to extend this analysis to systems with attractive interactions, such as the Lennard–Jones system, which features a well-defined energy minimum in the potential and exhibits more complex, temperature-dependent phase behavior~\cite{Massimo_2020PRLmelting,LJ_pd}.

\section*{Acknowledgments}
This work is supported by the National Natural Science Foundation of China (NSFC) (Grant Nos. 12422501 and 12374204).

\section*{Author contributions}
Y.Li conceived the project, Y.Li, R.Ding and W.Ma performed the research, Y.Li, R.Ding and W.Ma discussed the results. Y.Li wrote the paper.

\bibliographystyle{apsrev4-2}

\begin{thebibliography}{46}%
\makeatletter
\providecommand \@ifxundefined [1]{%
 \@ifx{#1\undefined}
}%
\providecommand \@ifnum [1]{%
 \ifnum #1\expandafter \@firstoftwo
 \else \expandafter \@secondoftwo
 \fi
}%
\providecommand \@ifx [1]{%
 \ifx #1\expandafter \@firstoftwo
 \else \expandafter \@secondoftwo
 \fi
}%
\providecommand \natexlab [1]{#1}%
\providecommand \enquote  [1]{``#1''}%
\providecommand \bibnamefont  [1]{#1}%
\providecommand \bibfnamefont [1]{#1}%
\providecommand \citenamefont [1]{#1}%
\providecommand \href@noop [0]{\@secondoftwo}%
\providecommand \href [0]{\begingroup \@sanitize@url \@href}%
\providecommand \@href[1]{\@@startlink{#1}\@@href}%
\providecommand \@@href[1]{\endgroup#1\@@endlink}%
\providecommand \@sanitize@url [0]{\catcode `\\12\catcode `\$12\catcode `\&12\catcode `\#12\catcode `\^12\catcode `\_12\catcode `\%12\relax}%
\providecommand \@@startlink[1]{}%
\providecommand \@@endlink[0]{}%
\providecommand \url  [0]{\begingroup\@sanitize@url \@url }%
\providecommand \@url [1]{\endgroup\@href {#1}{\urlprefix }}%
\providecommand \urlprefix  [0]{URL }%
\providecommand \Eprint [0]{\href }%
\providecommand \doibase [0]{https://doi.org/}%
\providecommand \selectlanguage [0]{\@gobble}%
\providecommand \bibinfo  [0]{\@secondoftwo}%
\providecommand \bibfield  [0]{\@secondoftwo}%
\providecommand \translation [1]{[#1]}%
\providecommand \BibitemOpen [0]{}%
\providecommand \bibitemStop [0]{}%
\providecommand \bibitemNoStop [0]{.\EOS\space}%
\providecommand \EOS [0]{\spacefactor3000\relax}%
\providecommand \BibitemShut  [1]{\csname bibitem#1\endcsname}%
\let\auto@bib@innerbib\@empty
\bibitem [{\citenamefont {Mermin}\ and\ \citenamefont {Wagner}(1966)}]{Mermin}%
  \BibitemOpen
  \bibfield  {author} {\bibinfo {author} {\bibfnamefont {N.~D.}\ \bibnamefont {Mermin}}\ and\ \bibinfo {author} {\bibfnamefont {H.}~\bibnamefont {Wagner}},\ }\href@noop {} {\bibfield  {journal} {\bibinfo  {journal} {Phys. Rev. Lett.}\ }\textbf {\bibinfo {volume} {17}},\ \bibinfo {pages} {1133} (\bibinfo {year} {1966})}\BibitemShut {NoStop}%
\bibitem [{\citenamefont {Kosterlitz}\ and\ \citenamefont {Thouless}(1973)}]{KT}%
  \BibitemOpen
  \bibfield  {author} {\bibinfo {author} {\bibfnamefont {J.~M.}\ \bibnamefont {Kosterlitz}}\ and\ \bibinfo {author} {\bibfnamefont {D.~J.}\ \bibnamefont {Thouless}},\ }\href@noop {} {\bibfield  {journal} {\bibinfo  {journal} {J. Phys. C}\ }\textbf {\bibinfo {volume} {6}},\ \bibinfo {pages} {1181} (\bibinfo {year} {1973})}\BibitemShut {NoStop}%
\bibitem [{\citenamefont {Halperin}\ and\ \citenamefont {Nelson}(1978)}]{HN}%
  \BibitemOpen
  \bibfield  {author} {\bibinfo {author} {\bibfnamefont {B.~I.}\ \bibnamefont {Halperin}}\ and\ \bibinfo {author} {\bibfnamefont {D.~R.}\ \bibnamefont {Nelson}},\ }\href@noop {} {\bibfield  {journal} {\bibinfo  {journal} {Phys. Rev. Lett.}\ }\textbf {\bibinfo {volume} {41}},\ \bibinfo {pages} {121} (\bibinfo {year} {1978})}\BibitemShut {NoStop}%
\bibitem [{\citenamefont {Young}(1979)}]{Y}%
  \BibitemOpen
  \bibfield  {author} {\bibinfo {author} {\bibfnamefont {A.~P.}\ \bibnamefont {Young}},\ }\href@noop {} {\bibfield  {journal} {\bibinfo  {journal} {Phys. Rev. B}\ }\textbf {\bibinfo {volume} {19}},\ \bibinfo {pages} {1855} (\bibinfo {year} {1979})}\BibitemShut {NoStop}%
\bibitem [{\citenamefont {Bernard}\ and\ \citenamefont {Krauth}(2011)}]{Krauth2011}%
  \BibitemOpen
  \bibfield  {author} {\bibinfo {author} {\bibfnamefont {E.~P.}\ \bibnamefont {Bernard}}\ and\ \bibinfo {author} {\bibfnamefont {W.}~\bibnamefont {Krauth}},\ }\href@noop {} {\bibfield  {journal} {\bibinfo  {journal} {Phys. Rev. Lett.}\ }\textbf {\bibinfo {volume} {107}},\ \bibinfo {pages} {155704} (\bibinfo {year} {2011})}\BibitemShut {NoStop}%
\bibitem [{\citenamefont {Thorneywork}\ \emph {et~al.}(2017)\citenamefont {Thorneywork}, \citenamefont {Abbott}, \citenamefont {Aarts},\ and\ \citenamefont {Dullens}}]{Experiment_harddisc}%
  \BibitemOpen
  \bibfield  {author} {\bibinfo {author} {\bibfnamefont {A.~L.}\ \bibnamefont {Thorneywork}}, \bibinfo {author} {\bibfnamefont {J.~L.}\ \bibnamefont {Abbott}}, \bibinfo {author} {\bibfnamefont {D.~G. A.~L.}\ \bibnamefont {Aarts}},\ and\ \bibinfo {author} {\bibfnamefont {R.~P.~A.}\ \bibnamefont {Dullens}},\ }\href@noop {} {\bibfield  {journal} {\bibinfo  {journal} {Phys. Rev. Lett.}\ }\textbf {\bibinfo {volume} {118}},\ \bibinfo {pages} {158001} (\bibinfo {year} {2017})}\BibitemShut {NoStop}%
\bibitem [{\citenamefont {Kapfer}\ and\ \citenamefont {Krauth}(2015)}]{Krauth2015}%
  \BibitemOpen
  \bibfield  {author} {\bibinfo {author} {\bibfnamefont {S.~C.}\ \bibnamefont {Kapfer}}\ and\ \bibinfo {author} {\bibfnamefont {W.}~\bibnamefont {Krauth}},\ }\href@noop {} {\bibfield  {journal} {\bibinfo  {journal} {Phys. Rev. Lett.}\ }\textbf {\bibinfo {volume} {114}},\ \bibinfo {pages} {035702} (\bibinfo {year} {2015})}\BibitemShut {NoStop}%
\bibitem [{\citenamefont {Zu}\ \emph {et~al.}(2016)\citenamefont {Zu}, \citenamefont {Liu}, \citenamefont {Tong},\ and\ \citenamefont {Xu}}]{ningxu}%
  \BibitemOpen
  \bibfield  {author} {\bibinfo {author} {\bibfnamefont {M.}~\bibnamefont {Zu}}, \bibinfo {author} {\bibfnamefont {J.}~\bibnamefont {Liu}}, \bibinfo {author} {\bibfnamefont {H.}~\bibnamefont {Tong}},\ and\ \bibinfo {author} {\bibfnamefont {N.}~\bibnamefont {Xu}},\ }\href@noop {} {\bibfield  {journal} {\bibinfo  {journal} {Phys. Rev. Lett.}\ }\textbf {\bibinfo {volume} {117}},\ \bibinfo {pages} {085702} (\bibinfo {year} {2016})}\BibitemShut {NoStop}%
\bibitem [{\citenamefont {Russo}\ and\ \citenamefont {Wilding}(2017)}]{John_Russo}%
  \BibitemOpen
  \bibfield  {author} {\bibinfo {author} {\bibfnamefont {J.}~\bibnamefont {Russo}}\ and\ \bibinfo {author} {\bibfnamefont {N.~B.}\ \bibnamefont {Wilding}},\ }\href@noop {} {\bibfield  {journal} {\bibinfo  {journal} {Phys. Rev. Lett.}\ }\textbf {\bibinfo {volume} {119}},\ \bibinfo {pages} {115702} (\bibinfo {year} {2017})}\BibitemShut {NoStop}%
\bibitem [{\citenamefont {Anderson}\ \emph {et~al.}(2017)\citenamefont {Anderson}, \citenamefont {Antonaglia}, \citenamefont {Millan}, \citenamefont {Engel},\ and\ \citenamefont {Glotzer}}]{Glotzer}%
  \BibitemOpen
  \bibfield  {author} {\bibinfo {author} {\bibfnamefont {J.~A.}\ \bibnamefont {Anderson}}, \bibinfo {author} {\bibfnamefont {J.}~\bibnamefont {Antonaglia}}, \bibinfo {author} {\bibfnamefont {J.~A.}\ \bibnamefont {Millan}}, \bibinfo {author} {\bibfnamefont {M.}~\bibnamefont {Engel}},\ and\ \bibinfo {author} {\bibfnamefont {S.~C.}\ \bibnamefont {Glotzer}},\ }\href@noop {} {\bibfield  {journal} {\bibinfo  {journal} {Phys. Rev. X}\ }\textbf {\bibinfo {volume} {7}},\ \bibinfo {pages} {021001} (\bibinfo {year} {2017})}\BibitemShut {NoStop}%
\bibitem [{\citenamefont {Komatsu}\ and\ \citenamefont {Tanaka}(2015)}]{Tanaka}%
  \BibitemOpen
  \bibfield  {author} {\bibinfo {author} {\bibfnamefont {Y.}~\bibnamefont {Komatsu}}\ and\ \bibinfo {author} {\bibfnamefont {H.}~\bibnamefont {Tanaka}},\ }\href@noop {} {\bibfield  {journal} {\bibinfo  {journal} {Phys. Rev. X}\ }\textbf {\bibinfo {volume} {5}},\ \bibinfo {pages} {031025} (\bibinfo {year} {2015})}\BibitemShut {NoStop}%
\bibitem [{\citenamefont {Li}\ and\ \citenamefont {Pica~Ciamarra}(2018)}]{Cell}%
  \BibitemOpen
  \bibfield  {author} {\bibinfo {author} {\bibfnamefont {Y.-W.}\ \bibnamefont {Li}}\ and\ \bibinfo {author} {\bibfnamefont {M.}~\bibnamefont {Pica~Ciamarra}},\ }\href@noop {} {\bibfield  {journal} {\bibinfo  {journal} {Phys. Rev. Mater.}\ }\textbf {\bibinfo {volume} {2}},\ \bibinfo {pages} {045602} (\bibinfo {year} {2018})}\BibitemShut {NoStop}%
\bibitem [{\citenamefont {Li}\ and\ \citenamefont {Ciamarra}(2020{\natexlab{a}})}]{LJ_pd}%
  \BibitemOpen
  \bibfield  {author} {\bibinfo {author} {\bibfnamefont {Y.-W.}\ \bibnamefont {Li}}\ and\ \bibinfo {author} {\bibfnamefont {M.~P.}\ \bibnamefont {Ciamarra}},\ }\href@noop {} {\bibfield  {journal} {\bibinfo  {journal} {Phys. Rev. E}\ }\textbf {\bibinfo {volume} {102}},\ \bibinfo {pages} {062101} (\bibinfo {year} {2020}{\natexlab{a}})}\BibitemShut {NoStop}%
\bibitem [{\citenamefont {Li}\ and\ \citenamefont {Ciamarra}(2020{\natexlab{b}})}]{Massimo_2020PRLmelting}%
  \BibitemOpen
  \bibfield  {author} {\bibinfo {author} {\bibfnamefont {Y.-W.}\ \bibnamefont {Li}}\ and\ \bibinfo {author} {\bibfnamefont {M.~P.}\ \bibnamefont {Ciamarra}},\ }\href@noop {} {\bibfield  {journal} {\bibinfo  {journal} {Phys. Rev. Lett.}\ }\textbf {\bibinfo {volume} {124}},\ \bibinfo {pages} {218002} (\bibinfo {year} {2020}{\natexlab{b}})}\BibitemShut {NoStop}%
\bibitem [{\citenamefont {Li}\ \emph {et~al.}(2023)\citenamefont {Li}, \citenamefont {Yao},\ and\ \citenamefont {Ciamarra}}]{YW_MP_prl2023}%
  \BibitemOpen
  \bibfield  {author} {\bibinfo {author} {\bibfnamefont {Y.-W.}\ \bibnamefont {Li}}, \bibinfo {author} {\bibfnamefont {Y.}~\bibnamefont {Yao}},\ and\ \bibinfo {author} {\bibfnamefont {M.~P.}\ \bibnamefont {Ciamarra}},\ }\href@noop {} {\bibfield  {journal} {\bibinfo  {journal} {Phys. Rev. Lett.}\ }\textbf {\bibinfo {volume} {130}},\ \bibinfo {pages} {258202} (\bibinfo {year} {2023})}\BibitemShut {NoStop}%
\bibitem [{\citenamefont {Zhu}\ and\ \citenamefont {Wang}(2024)}]{zhuwang_NC2024}%
  \BibitemOpen
  \bibfield  {author} {\bibinfo {author} {\bibfnamefont {R.}~\bibnamefont {Zhu}}\ and\ \bibinfo {author} {\bibfnamefont {Y.}~\bibnamefont {Wang}},\ }\href@noop {} {\bibfield  {journal} {\bibinfo  {journal} {Nat. Commun.}\ }\textbf {\bibinfo {volume} {15}},\ \bibinfo {pages} {6389} (\bibinfo {year} {2024})}\BibitemShut {NoStop}%
\bibitem [{\citenamefont {Li}\ \emph {et~al.}(2019)\citenamefont {Li}, \citenamefont {Xiao}, \citenamefont {Wang}, \citenamefont {Wen},\ and\ \citenamefont {Wang}}]{ZirenPRX}%
  \BibitemOpen
  \bibfield  {author} {\bibinfo {author} {\bibfnamefont {B.}~\bibnamefont {Li}}, \bibinfo {author} {\bibfnamefont {X.}~\bibnamefont {Xiao}}, \bibinfo {author} {\bibfnamefont {S.}~\bibnamefont {Wang}}, \bibinfo {author} {\bibfnamefont {W.}~\bibnamefont {Wen}},\ and\ \bibinfo {author} {\bibfnamefont {Z.}~\bibnamefont {Wang}},\ }\href@noop {} {\bibfield  {journal} {\bibinfo  {journal} {Phys. Rev. X}\ }\textbf {\bibinfo {volume} {9}},\ \bibinfo {pages} {031032} (\bibinfo {year} {2019})}\BibitemShut {NoStop}%
\bibitem [{\citenamefont {Loewe}\ \emph {et~al.}(2020)\citenamefont {Loewe}, \citenamefont {Chiang}, \citenamefont {Marenduzzo},\ and\ \citenamefont {Marchetti}}]{deform_PRL2020}%
  \BibitemOpen
  \bibfield  {author} {\bibinfo {author} {\bibfnamefont {B.}~\bibnamefont {Loewe}}, \bibinfo {author} {\bibfnamefont {M.}~\bibnamefont {Chiang}}, \bibinfo {author} {\bibfnamefont {D.}~\bibnamefont {Marenduzzo}},\ and\ \bibinfo {author} {\bibfnamefont {M.~C.}\ \bibnamefont {Marchetti}},\ }\href@noop {} {\bibfield  {journal} {\bibinfo  {journal} {Phys. Rev. Lett.}\ }\textbf {\bibinfo {volume} {125}},\ \bibinfo {pages} {038003} (\bibinfo {year} {2020})}\BibitemShut {NoStop}%
\bibitem [{\citenamefont {Du}\ \emph {et~al.}(2017)\citenamefont {Du}, \citenamefont {Doxastakis}, \citenamefont {Hilou},\ and\ \citenamefont {Biswal}}]{dudi_softmatter}%
  \BibitemOpen
  \bibfield  {author} {\bibinfo {author} {\bibfnamefont {D.}~\bibnamefont {Du}}, \bibinfo {author} {\bibfnamefont {M.}~\bibnamefont {Doxastakis}}, \bibinfo {author} {\bibfnamefont {E.}~\bibnamefont {Hilou}},\ and\ \bibinfo {author} {\bibfnamefont {S.~L.}\ \bibnamefont {Biswal}},\ }\href@noop {} {\bibfield  {journal} {\bibinfo  {journal} {Soft Matter}\ }\textbf {\bibinfo {volume} {13}},\ \bibinfo {pages} {1548} (\bibinfo {year} {2017})}\BibitemShut {NoStop}%
\bibitem [{\citenamefont {Sampedro~Ruiz}\ \emph {et~al.}(2019)\citenamefont {Sampedro~Ruiz}, \citenamefont {Lei},\ and\ \citenamefont {Ni}}]{Ran2dmelt}%
  \BibitemOpen
  \bibfield  {author} {\bibinfo {author} {\bibfnamefont {P.}~\bibnamefont {Sampedro~Ruiz}}, \bibinfo {author} {\bibfnamefont {Q.-l.}\ \bibnamefont {Lei}},\ and\ \bibinfo {author} {\bibfnamefont {R.}~\bibnamefont {Ni}},\ }\href@noop {} {\bibfield  {journal} {\bibinfo  {journal} {Commun. Phys.}\ }\textbf {\bibinfo {volume} {2}},\ \bibinfo {pages} {70} (\bibinfo {year} {2019})}\BibitemShut {NoStop}%
\bibitem [{\citenamefont {Guo}\ \emph {et~al.}(2021)\citenamefont {Guo}, \citenamefont {Nie},\ and\ \citenamefont {Xu}}]{Guo2021}%
  \BibitemOpen
  \bibfield  {author} {\bibinfo {author} {\bibfnamefont {J.}~\bibnamefont {Guo}}, \bibinfo {author} {\bibfnamefont {Y.}~\bibnamefont {Nie}},\ and\ \bibinfo {author} {\bibfnamefont {N.}~\bibnamefont {Xu}},\ }\href@noop {} {\bibfield  {journal} {\bibinfo  {journal} {Soft Matter}\ }\textbf {\bibinfo {volume} {17}},\ \bibinfo {pages} {3397} (\bibinfo {year} {2021})}\BibitemShut {NoStop}%
\bibitem [{\citenamefont {Klamser}\ \emph {et~al.}(2018)\citenamefont {Klamser}, \citenamefont {Kapfer},\ and\ \citenamefont {Krauth}}]{Krauth_NC2018}%
  \BibitemOpen
  \bibfield  {author} {\bibinfo {author} {\bibfnamefont {J.~U.}\ \bibnamefont {Klamser}}, \bibinfo {author} {\bibfnamefont {S.~C.}\ \bibnamefont {Kapfer}},\ and\ \bibinfo {author} {\bibfnamefont {W.}~\bibnamefont {Krauth}},\ }\href@noop {} {\bibfield  {journal} {\bibinfo  {journal} {Nat. Commun.}\ }\textbf {\bibinfo {volume} {9}},\ \bibinfo {pages} {5045} (\bibinfo {year} {2018})}\BibitemShut {NoStop}%
\bibitem [{\citenamefont {Deutschl\"{a}nder}\ \emph {et~al.}(2013)\citenamefont {Deutschl\"{a}nder}, \citenamefont {Horn}, \citenamefont {L\"{o}wen}, \citenamefont {Maret},\ and\ \citenamefont {Keim}}]{pinned_particle}%
  \BibitemOpen
  \bibfield  {author} {\bibinfo {author} {\bibfnamefont {S.}~\bibnamefont {Deutschl\"{a}nder}}, \bibinfo {author} {\bibfnamefont {T.}~\bibnamefont {Horn}}, \bibinfo {author} {\bibfnamefont {H.}~\bibnamefont {L\"{o}wen}}, \bibinfo {author} {\bibfnamefont {G.}~\bibnamefont {Maret}},\ and\ \bibinfo {author} {\bibfnamefont {P.}~\bibnamefont {Keim}},\ }\href@noop {} {\bibfield  {journal} {\bibinfo  {journal} {Phys. Rev. Lett.}\ }\textbf {\bibinfo {volume} {111}},\ \bibinfo {pages} {098301} (\bibinfo {year} {2013})}\BibitemShut {NoStop}%
\bibitem [{\citenamefont {Qi}\ and\ \citenamefont {Dijkstra}(2015)}]{qi_sm2015}%
  \BibitemOpen
  \bibfield  {author} {\bibinfo {author} {\bibfnamefont {W.}~\bibnamefont {Qi}}\ and\ \bibinfo {author} {\bibfnamefont {M.}~\bibnamefont {Dijkstra}},\ }\href@noop {} {\bibfield  {journal} {\bibinfo  {journal} {Soft Matter}\ }\textbf {\bibinfo {volume} {11}},\ \bibinfo {pages} {2852} (\bibinfo {year} {2015})}\BibitemShut {NoStop}%
\bibitem [{\citenamefont {Toledano}\ \emph {et~al.}(2021)\citenamefont {Toledano}, \citenamefont {Pancorbo}, \citenamefont {Alvarellos},\ and\ \citenamefont {G\'{a}lvez}}]{Toledano_PRB}%
  \BibitemOpen
  \bibfield  {author} {\bibinfo {author} {\bibfnamefont {O.}~\bibnamefont {Toledano}}, \bibinfo {author} {\bibfnamefont {M.}~\bibnamefont {Pancorbo}}, \bibinfo {author} {\bibfnamefont {J.~E.}\ \bibnamefont {Alvarellos}},\ and\ \bibinfo {author} {\bibfnamefont {O.}~\bibnamefont {G\'{a}lvez}},\ }\href@noop {} {\bibfield  {journal} {\bibinfo  {journal} {Phys. Rev. B}\ }\textbf {\bibinfo {volume} {103}},\ \bibinfo {pages} {094107} (\bibinfo {year} {2021})}\BibitemShut {NoStop}%
\bibitem [{\citenamefont {Digregorio}\ \emph {et~al.}(2018)\citenamefont {Digregorio}, \citenamefont {Levis}, \citenamefont {Suma}, \citenamefont {Cugliandolo}, \citenamefont {Gonnella},\ and\ \citenamefont {Pagonabarraga}}]{activemelting_prl}%
  \BibitemOpen
  \bibfield  {author} {\bibinfo {author} {\bibfnamefont {P.}~\bibnamefont {Digregorio}}, \bibinfo {author} {\bibfnamefont {D.}~\bibnamefont {Levis}}, \bibinfo {author} {\bibfnamefont {A.}~\bibnamefont {Suma}}, \bibinfo {author} {\bibfnamefont {L.~F.}\ \bibnamefont {Cugliandolo}}, \bibinfo {author} {\bibfnamefont {G.}~\bibnamefont {Gonnella}},\ and\ \bibinfo {author} {\bibfnamefont {I.}~\bibnamefont {Pagonabarraga}},\ }\href@noop {} {\bibfield  {journal} {\bibinfo  {journal} {Phys. Rev. Lett.}\ }\textbf {\bibinfo {volume} {121}},\ \bibinfo {pages} {098003} (\bibinfo {year} {2018})}\BibitemShut {NoStop}%
\bibitem [{\citenamefont {Cugliandolo}\ \emph {et~al.}(2017)\citenamefont {Cugliandolo}, \citenamefont {Digregorio}, \citenamefont {Gonnella},\ and\ \citenamefont {Suma}}]{Cugliandolo_prl}%
  \BibitemOpen
  \bibfield  {author} {\bibinfo {author} {\bibfnamefont {L.~F.}\ \bibnamefont {Cugliandolo}}, \bibinfo {author} {\bibfnamefont {P.}~\bibnamefont {Digregorio}}, \bibinfo {author} {\bibfnamefont {G.}~\bibnamefont {Gonnella}},\ and\ \bibinfo {author} {\bibfnamefont {A.}~\bibnamefont {Suma}},\ }\href@noop {} {\bibfield  {journal} {\bibinfo  {journal} {Phys. Rev. Lett.}\ }\textbf {\bibinfo {volume} {119}},\ \bibinfo {pages} {268002} (\bibinfo {year} {2017})}\BibitemShut {NoStop}%
\bibitem [{\citenamefont {Bui}\ \emph {et~al.}(2025)\citenamefont {Bui}, \citenamefont {Lamprecht}, \citenamefont {Madsen}, \citenamefont {Kurpas}, \citenamefont {Kotrusz}, \citenamefont {Markevich}, \citenamefont {Mangler}, \citenamefont {Kotakoski}, \citenamefont {Filipovic}, \citenamefont {Meyer}, \citenamefont {Pennycook}, \citenamefont {Skákalová},\ and\ \citenamefont {Mustonen}}]{Bui_Science}%
  \BibitemOpen
  \bibfield  {author} {\bibinfo {author} {\bibfnamefont {T.~A.}\ \bibnamefont {Bui}}, \bibinfo {author} {\bibfnamefont {D.}~\bibnamefont {Lamprecht}}, \bibinfo {author} {\bibfnamefont {J.}~\bibnamefont {Madsen}}, \bibinfo {author} {\bibfnamefont {M.}~\bibnamefont {Kurpas}}, \bibinfo {author} {\bibfnamefont {P.}~\bibnamefont {Kotrusz}}, \bibinfo {author} {\bibfnamefont {A.}~\bibnamefont {Markevich}}, \bibinfo {author} {\bibfnamefont {C.}~\bibnamefont {Mangler}}, \bibinfo {author} {\bibfnamefont {J.}~\bibnamefont {Kotakoski}}, \bibinfo {author} {\bibfnamefont {L.}~\bibnamefont {Filipovic}}, \bibinfo {author} {\bibfnamefont {J.~C.}\ \bibnamefont {Meyer}}, \bibinfo {author} {\bibfnamefont {T.~J.}\ \bibnamefont {Pennycook}}, \bibinfo {author} {\bibfnamefont {V.}~\bibnamefont {Skákalová}},\ and\ \bibinfo {author} {\bibfnamefont {K.}~\bibnamefont {Mustonen}},\ }\href {https://doi.org/doi:10.1126/science.adv7915} {\bibfield  {journal} {\bibinfo  {journal} {Science}\ }\textbf {\bibinfo {volume} {390}},\ \bibinfo
  {pages} {1033} (\bibinfo {year} {2025})}\BibitemShut {NoStop}%
\bibitem [{\citenamefont {Prestipino}\ \emph {et~al.}(2005)\citenamefont {Prestipino}, \citenamefont {Saija},\ and\ \citenamefont {Giaquinta}}]{Gauss3d}%
  \BibitemOpen
  \bibfield  {author} {\bibinfo {author} {\bibfnamefont {S.}~\bibnamefont {Prestipino}}, \bibinfo {author} {\bibfnamefont {F.}~\bibnamefont {Saija}},\ and\ \bibinfo {author} {\bibfnamefont {P.~V.}\ \bibnamefont {Giaquinta}},\ }\href@noop {} {\bibfield  {journal} {\bibinfo  {journal} {Phys. Rev. E}\ }\textbf {\bibinfo {volume} {71}},\ \bibinfo {pages} {050102} (\bibinfo {year} {2005})}\BibitemShut {NoStop}%
\bibitem [{\citenamefont {Allen}(1987)}]{Allen_book}%
  \BibitemOpen
  \bibfield  {author} {\bibinfo {author} {\bibfnamefont {M.}~\bibnamefont {Allen}},\ }\href@noop {} {\emph {\bibinfo {title} {Computer Simulation of Liquids}}}\ (\bibinfo  {publisher} {Oxford University Press, Oxford},\ \bibinfo {year} {1987})\BibitemShut {NoStop}%
\bibitem [{\citenamefont {Zhu}\ \emph {et~al.}(2013)\citenamefont {Zhu}, \citenamefont {Liu}, \citenamefont {Li}, \citenamefont {Qian}, \citenamefont {Milano},\ and\ \citenamefont {Lu}}]{Galamost}%
  \BibitemOpen
  \bibfield  {author} {\bibinfo {author} {\bibfnamefont {Y.}~\bibnamefont {Zhu}}, \bibinfo {author} {\bibfnamefont {H.}~\bibnamefont {Liu}}, \bibinfo {author} {\bibfnamefont {Z.}~\bibnamefont {Li}}, \bibinfo {author} {\bibfnamefont {H.}~\bibnamefont {Qian}}, \bibinfo {author} {\bibfnamefont {G.}~\bibnamefont {Milano}},\ and\ \bibinfo {author} {\bibfnamefont {Z.}~\bibnamefont {Lu}},\ }\href@noop {} {\bibfield  {journal} {\bibinfo  {journal} {J. Comput. Chem.}\ }\textbf {\bibinfo {volume} {34}},\ \bibinfo {pages} {2197} (\bibinfo {year} {2013})}\BibitemShut {NoStop}%
\bibitem [{\citenamefont {Plimpton}(1995)}]{Lammps}%
  \BibitemOpen
  \bibfield  {author} {\bibinfo {author} {\bibfnamefont {S.}~\bibnamefont {Plimpton}},\ }\href@noop {} {\bibfield  {journal} {\bibinfo  {journal} {J. Comput. Phys.}\ }\textbf {\bibinfo {volume} {117}},\ \bibinfo {pages} {1} (\bibinfo {year} {1995})}\BibitemShut {NoStop}%
\bibitem [{\citenamefont {Shintani}\ and\ \citenamefont {Tanaka}(2008)}]{Shintani_NM}%
  \BibitemOpen
  \bibfield  {author} {\bibinfo {author} {\bibfnamefont {H.}~\bibnamefont {Shintani}}\ and\ \bibinfo {author} {\bibfnamefont {H.}~\bibnamefont {Tanaka}},\ }\href {https://doi.org/10.1038/nmat2293} {\bibfield  {journal} {\bibinfo  {journal} {Nat. Mater.}\ }\textbf {\bibinfo {volume} {7}},\ \bibinfo {pages} {870} (\bibinfo {year} {2008})}\BibitemShut {NoStop}%
\bibitem [{\citenamefont {Hu}\ and\ \citenamefont {Tanaka}(2022)}]{Hu_NP}%
  \BibitemOpen
  \bibfield  {author} {\bibinfo {author} {\bibfnamefont {Y.-C.}\ \bibnamefont {Hu}}\ and\ \bibinfo {author} {\bibfnamefont {H.}~\bibnamefont {Tanaka}},\ }\href {https://doi.org/10.1038/s41567-022-01628-6} {\bibfield  {journal} {\bibinfo  {journal} {Nat. Phys.}\ }\textbf {\bibinfo {volume} {18}},\ \bibinfo {pages} {669} (\bibinfo {year} {2022})}\BibitemShut {NoStop}%
\bibitem [{\citenamefont {Smith}\ \emph {et~al.}()\citenamefont {Smith}, \citenamefont {Li}, \citenamefont {Hoff}, \citenamefont {Garrett}, \citenamefont {Kim}, \citenamefont {Yang}, \citenamefont {Lucas}, \citenamefont {Swan-Wood}, \citenamefont {Lin}, \citenamefont {Stone}, \citenamefont {Abernathy}, \citenamefont {Demetriou},\ and\ \citenamefont {Fultz}}]{Smith_NP}%
  \BibitemOpen
  \bibfield  {author} {\bibinfo {author} {\bibfnamefont {H.~L.}\ \bibnamefont {Smith}}, \bibinfo {author} {\bibfnamefont {C.~W.}\ \bibnamefont {Li}}, \bibinfo {author} {\bibfnamefont {A.}~\bibnamefont {Hoff}}, \bibinfo {author} {\bibfnamefont {G.~R.}\ \bibnamefont {Garrett}}, \bibinfo {author} {\bibfnamefont {D.~S.}\ \bibnamefont {Kim}}, \bibinfo {author} {\bibfnamefont {F.~C.}\ \bibnamefont {Yang}}, \bibinfo {author} {\bibfnamefont {M.~S.}\ \bibnamefont {Lucas}}, \bibinfo {author} {\bibfnamefont {T.}~\bibnamefont {Swan-Wood}}, \bibinfo {author} {\bibfnamefont {J.~Y.~Y.}\ \bibnamefont {Lin}}, \bibinfo {author} {\bibfnamefont {M.~B.}\ \bibnamefont {Stone}}, \bibinfo {author} {\bibfnamefont {D.~L.}\ \bibnamefont {Abernathy}}, \bibinfo {author} {\bibfnamefont {M.~D.}\ \bibnamefont {Demetriou}},\ and\ \bibinfo {author} {\bibfnamefont {B.}~\bibnamefont {Fultz}},\ }\href {https://doi.org/10.1038/nphys4142} {\bibfield  {journal} {\bibinfo  {journal} {Nat. Phys.}\ }\textbf {\bibinfo {volume} {13}},\ \bibinfo {pages}
  {900}}\BibitemShut {NoStop}%
\bibitem [{\citenamefont {Allen}(2015)}]{Allen_PRB}%
  \BibitemOpen
  \bibfield  {author} {\bibinfo {author} {\bibfnamefont {P.~B.}\ \bibnamefont {Allen}},\ }\href {https://doi.org/10.1103/PhysRevB.92.064106} {\bibfield  {journal} {\bibinfo  {journal} {Phys. Rev. B}\ }\textbf {\bibinfo {volume} {92}},\ \bibinfo {pages} {064106} (\bibinfo {year} {2015})}\BibitemShut {NoStop}%
\bibitem [{\citenamefont {Mayer}\ and\ \citenamefont {Wood}(1965)}]{Mayer_wood}%
  \BibitemOpen
  \bibfield  {author} {\bibinfo {author} {\bibfnamefont {J.~E.}\ \bibnamefont {Mayer}}\ and\ \bibinfo {author} {\bibfnamefont {W.~W.}\ \bibnamefont {Wood}},\ }\href@noop {} {\bibfield  {journal} {\bibinfo  {journal} {J. Chem. Phys.}\ }\textbf {\bibinfo {volume} {42}},\ \bibinfo {pages} {4268} (\bibinfo {year} {1965})}\BibitemShut {NoStop}%
\bibitem [{\citenamefont {Sciortino}(2005)}]{Sciortino_2005}%
  \BibitemOpen
  \bibfield  {author} {\bibinfo {author} {\bibfnamefont {F.}~\bibnamefont {Sciortino}},\ }\href {https://doi.org/10.1088/1742-5468/2005/05/P05015} {\bibfield  {journal} {\bibinfo  {journal} {J. Stat. Mech.}\ }\textbf {\bibinfo {volume} {2005}},\ \bibinfo {pages} {P05015} (\bibinfo {year} {2005})}\BibitemShut {NoStop}%
\bibitem [{\citenamefont {Giovambattista}\ \emph {et~al.}(2016)\citenamefont {Giovambattista}, \citenamefont {Sciortino}, \citenamefont {Starr},\ and\ \citenamefont {Poole}}]{Sciortino2016}%
  \BibitemOpen
  \bibfield  {author} {\bibinfo {author} {\bibfnamefont {N.}~\bibnamefont {Giovambattista}}, \bibinfo {author} {\bibfnamefont {F.}~\bibnamefont {Sciortino}}, \bibinfo {author} {\bibfnamefont {F.~W.}\ \bibnamefont {Starr}},\ and\ \bibinfo {author} {\bibfnamefont {P.~H.}\ \bibnamefont {Poole}},\ }\href {https://doi.org/10.1063/1.4968047} {\bibfield  {journal} {\bibinfo  {journal} {J. Chem. Phys.}\ }\textbf {\bibinfo {volume} {145}},\ \bibinfo {pages} {224501} (\bibinfo {year} {2016})}\BibitemShut {NoStop}%
\bibitem [{\citenamefont {Sun}\ \emph {et~al.}(2018)\citenamefont {Sun}, \citenamefont {Xu},\ and\ \citenamefont {Giovambattista}}]{gangsunPRL2018}%
  \BibitemOpen
  \bibfield  {author} {\bibinfo {author} {\bibfnamefont {G.}~\bibnamefont {Sun}}, \bibinfo {author} {\bibfnamefont {L.}~\bibnamefont {Xu}},\ and\ \bibinfo {author} {\bibfnamefont {N.}~\bibnamefont {Giovambattista}},\ }\href {https://doi.org/10.1103/PhysRevLett.120.035701} {\bibfield  {journal} {\bibinfo  {journal} {Phys. Rev. Lett.}\ }\textbf {\bibinfo {volume} {120}},\ \bibinfo {pages} {035701} (\bibinfo {year} {2018})}\BibitemShut {NoStop}%
\bibitem [{\citenamefont {Bitzek}\ \emph {et~al.}(2006)\citenamefont {Bitzek}, \citenamefont {Koskinen}, \citenamefont {G\"ahler}, \citenamefont {Moseler},\ and\ \citenamefont {Gumbsch}}]{Fire}%
  \BibitemOpen
  \bibfield  {author} {\bibinfo {author} {\bibfnamefont {E.}~\bibnamefont {Bitzek}}, \bibinfo {author} {\bibfnamefont {P.}~\bibnamefont {Koskinen}}, \bibinfo {author} {\bibfnamefont {F.}~\bibnamefont {G\"ahler}}, \bibinfo {author} {\bibfnamefont {M.}~\bibnamefont {Moseler}},\ and\ \bibinfo {author} {\bibfnamefont {P.}~\bibnamefont {Gumbsch}},\ }\href@noop {} {\bibfield  {journal} {\bibinfo  {journal} {Phys. Rev. Lett.}\ }\textbf {\bibinfo {volume} {97}},\ \bibinfo {pages} {170201} (\bibinfo {year} {2006})}\BibitemShut {NoStop}%
\bibitem [{\citenamefont {Ma}\ and\ \citenamefont {Li}(2025)}]{MaPRE}%
  \BibitemOpen
  \bibfield  {author} {\bibinfo {author} {\bibfnamefont {W.-H.}\ \bibnamefont {Ma}}\ and\ \bibinfo {author} {\bibfnamefont {Y.-W.}\ \bibnamefont {Li}},\ }\href {https://doi.org/10.1103/t66b-drh4} {\bibfield  {journal} {\bibinfo  {journal} {Phys. Rev. E}\ }\textbf {\bibinfo {volume} {112}},\ \bibinfo {pages} {034117} (\bibinfo {year} {2025})}\BibitemShut {NoStop}%
\bibitem [{\citenamefont {Huang}\ and\ \citenamefont {Widom}(2022)}]{Huang}%
  \BibitemOpen
  \bibfield  {author} {\bibinfo {author} {\bibfnamefont {Y.}~\bibnamefont {Huang}}\ and\ \bibinfo {author} {\bibfnamefont {M.}~\bibnamefont {Widom}},\ }\bibfield  {journal} {\bibinfo  {journal} {Entropy}\ }\textbf {\bibinfo {volume} {24}},\ \href {https://doi.org/10.3390/e24050618} {10.3390/e24050618} (\bibinfo {year} {2022})\BibitemShut {NoStop}%
\bibitem [{\citenamefont {Johari}(2000)}]{Johari2000}%
  \BibitemOpen
  \bibfield  {author} {\bibinfo {author} {\bibfnamefont {G.~P.}\ \bibnamefont {Johari}},\ }\href {https://doi.org/10.1063/1.481349} {\bibfield  {journal} {\bibinfo  {journal} {J. Chem. Phys.}\ }\textbf {\bibinfo {volume} {112}},\ \bibinfo {pages} {7518} (\bibinfo {year} {2000})}\BibitemShut {NoStop}%
\bibitem [{\citenamefont {Digregorio}\ \emph {et~al.}(2022)\citenamefont {Digregorio}, \citenamefont {Levis}, \citenamefont {Cugliandolo}, \citenamefont {Gonnella},\ and\ \citenamefont {Pagonabarraga}}]{digregorio2019clustering}%
  \BibitemOpen
  \bibfield  {author} {\bibinfo {author} {\bibfnamefont {P.}~\bibnamefont {Digregorio}}, \bibinfo {author} {\bibfnamefont {D.}~\bibnamefont {Levis}}, \bibinfo {author} {\bibfnamefont {L.~F.}\ \bibnamefont {Cugliandolo}}, \bibinfo {author} {\bibfnamefont {G.}~\bibnamefont {Gonnella}},\ and\ \bibinfo {author} {\bibfnamefont {I.}~\bibnamefont {Pagonabarraga}},\ }\href {https://doi.org/10.1039/D1SM01411K} {\bibfield  {journal} {\bibinfo  {journal} {Soft Matter}\ }\textbf {\bibinfo {volume} {18}},\ \bibinfo {pages} {566} (\bibinfo {year} {2022})}\BibitemShut {NoStop}%
\bibitem [{\citenamefont {Pàmies}\ \emph {et~al.}(2009)\citenamefont {Pàmies}, \citenamefont {Cacciuto},\ and\ \citenamefont {Frenkel}}]{hert3d}%
  \BibitemOpen
  \bibfield  {author} {\bibinfo {author} {\bibfnamefont {J.~C.}\ \bibnamefont {Pàmies}}, \bibinfo {author} {\bibfnamefont {A.}~\bibnamefont {Cacciuto}},\ and\ \bibinfo {author} {\bibfnamefont {D.}~\bibnamefont {Frenkel}},\ }\href@noop {} {\bibfield  {journal} {\bibinfo  {journal} {J. Chem. Phys.}\ }\textbf {\bibinfo {volume} {131}},\ \bibinfo {pages} {044514} (\bibinfo {year} {2009})}\BibitemShut {NoStop}%
\end{thebibliography}
%
\end{document}